\documentclass[%
reprint,
superscriptaddress,
 amsmath,amssymb,
 aps,
 prb,
]{revtex4-1}

\newcommand*{\degc}{$\!^\circ$C}
\newcommand*{\dego}{$\!^\circ$}
\newcommand*{\wn}{$\mathrm{cm^{-1}}$}

\usepackage{graphicx}% Include figure files
\usepackage{dcolumn}% Align table columns on decimal point
\usepackage{bm}% bold math
\usepackage{makecell}
\usepackage{multirow}
\begin{document}

\preprint{APS/123-QED}

\title{Electronic and vibrational signatures of ruthenium vacancies in Sr$_2$RuO$_4$ thin films}% Force line breaks with \\

\author{Gideok Kim}
\affiliation{Max-Planck-Institute for Solid State Research, Heisenbergstrasse 1, 70569 Stuttgart, Germany}%
\author{Y. Eren Suyolcu}%
\affiliation{Max-Planck-Institute for Solid State Research, Heisenbergstrasse 1, 70569 Stuttgart, Germany}%
\author{J. Herrero-Martin}%
\affiliation{ALBA Synchrotron Light Source, E-08290 Cerdanyola del vall\'es, Barcelona, Spain
}%
\author{D. Putzky}
\affiliation{Max-Planck-Institute for Solid State Research, Heisenbergstrasse 1, 70569 Stuttgart, Germany}%
\author{H. P. Nair}%
\affiliation{Department of Materials Science and Engineering, Cornell University, Ithaca, New York 14853, USA
}%
\author{J. P. Ruf}%
\affiliation{Department of Physics, Cornell University, Ithaca, New York 14853, USA
}%
\author{N. J. Schreiber}%
\affiliation{Department of Materials Science and Engineering, Cornell University, Ithaca, New York 14853, USA
}%
\author{C. Dietl}%
\affiliation{Max-Planck-Institute for Solid State Research, Heisenbergstrasse 1, 70569 Stuttgart, Germany}%
\affiliation{Center for Artificial Low Dimensional Electronic Systems, Institute for Basic Science (IBS), 77 Cheongam-Ro, Pohang 790-784, South Korea
}
\affiliation{Advanced Photon Source, Argonne National Laboratory, Lemont, Illinois 60439, USA
}%
\author{G. Christiani}%
\affiliation{Max-Planck-Institute for Solid State Research, Heisenbergstrasse 1, 70569 Stuttgart, Germany}%
\author{G. Logvenov}%
\affiliation{Max-Planck-Institute for Solid State Research, Heisenbergstrasse 1, 70569 Stuttgart, Germany}%
\author{M. Minola}%
\affiliation{Max-Planck-Institute for Solid State Research, Heisenbergstrasse 1, 70569 Stuttgart, Germany}%
\author{P. A. van Aken}%
\affiliation{Max-Planck-Institute for Solid State Research, Heisenbergstrasse 1, 70569 Stuttgart, Germany}%
\author{K. M. Shen}%
\affiliation{Department of Physics, Cornell University, Ithaca, New York 14853, USA
}%
\affiliation{Kavli Institute at Cornell for Nanoscale Science, Ithaca, New York 14853, USA
}%
\author{D. G. Schlom}%
\affiliation{Department of Materials Science and Engineering, Cornell University, Ithaca, New York 14853, USA
}%
\affiliation{Kavli Institute at Cornell for Nanoscale Science, Ithaca, New York 14853, USA
}%
\author{B. Keimer}%
\email{B.Keimer@fkf.mpg.de}
\affiliation{Max-Planck-Institute for Solid State Research, Heisenbergstrasse 1, 70569 Stuttgart, Germany}%

\date{\today}% It is always \today, today,

\begin{abstract}
The synthesis of stoichiometric Sr$_2$RuO$_4$ thin films has been a challenge because of the high volatility of ruthenium oxide precursors, which gives rise to ruthenium vacancies in the films.
Ru vacancies greatly affect the transport properties and electronic phase behavior of Sr$_2$RuO$_4$, but their direct detection is difficult due to their atomic dimensions and low concentration.
We applied polarized X-ray absorption spectroscopy at the oxygen K-edge and confocal Raman spectroscopy to Sr$_2$RuO$_4$ thin films synthesized under different conditions. The results show that these methods can serve as sensitive probes of the electronic and vibrational properties of Ru vacancies, respectively.
The intensities of the vacancy-related spectroscopic features extracted from these measurements are well correlated with the transport properties of the films.
The methodology introduced here can thus help to understand and control the stoichiometry and transport properties in films of Sr$_2$RuO$_4$ and other ruthenates. 

\end{abstract}

\maketitle

%\tableofcontents

%%%%%%%%%%%%%%%%%%%%%%%% paragraph Introduction %%%%%%%%%%%%%%%%%%%%%%%%%%
\section{Introduction}
Ruthenium oxides have long served as model compounds for the influence of spin-orbit interactions on the electronic properties of strongly correlated electron systems.
The layered compound Sr$_2$RuO$_4$ has attracted particular attention because it exhibits textbook Fermi liquid behavior as well as an unconventional superconducting state whose microscopic description continues to be strongly debated \cite{Mackenzie1996,Mackenzie2017,Pustogow2019}.
As the valence electrons reside in the nearly degenerate $t_{2g}$ levels of the Ru ions in the tetragonal crystal field, externally imposed lattice distortions in the form of uniaxial \cite{Steppke2017,Barber2018} or biaxial strain \cite{Burganov2016,Hsu2016} have been shown to profoundly affect the phase behavior and physical properties.
In particular, recent angle-resolved photoelectron spectroscopy studies performed \textit{in-situ} on Sr$_2$RuO$_4$ thin films demonstrated that the Fermi surface is very sensitive to epitaxial strain \cite{Burganov2016,Hsu2016}. 

Since its in-plane lattice parameters (\textit{a}=\textit{b}=3.87\AA) are similar to other ternary transition metal oxides that exhibit novel physical properties, Sr$_2$RuO$_4$ can be readily integrated into all-oxide thin film devices.
Compared to most functional transition metal oxides, which are insulators, several ruthenates including Sr$_2$RuO$_4$ and its cubic perovskite analogue SrRuO$_3$ exhibit good metallic properties, which make them suitable as electrode materials for oxide electronics \cite{Koster2012}. 
In several recent studies, for example, SrRuO$_3$ is used as a standard electrode for ferroelectric tunnel junctions \cite{Garcia2014}.
Sr$_2$RuO$_4$, on the other hand, has high thermal stability even up to 1000 \degc, which makes it an appealing candidate for a bottom electrode that must withstand the high growth temperatures of oxides deposited on top of it \cite{Takahasi2017}.
In addition, SrRuO$_3$-Sr$_2$RuO$_4$ heterostructures have recently been used to explore fundamental properties of the superconducting state in Sr$_2$RuO$_4$ \cite{Anwar2016,Chung2018,Mufazalova2018}.

Despite the fundamental interest and potential applications, the growth of high quality ruthenate thin films has proven to be very challenging due to the nature of ruthenium and its oxides \cite{Nair2018-113,Nair2018-214,Marshall2017,Cao2016,Marshall2017,Uchida2017,Takahasi2017,Krockenberger2010}. 
In fact, the high volatility of ruthenium oxides (RuO$_{x = 2, 3, 4}$) leads to ruthenium deficiency, as shown in numerous reports \cite{Siemons2007,Dabrowski2004,Nair2018-113,Nair2018-214,Schraknepper2015,Schraknepper2016}. 
The ruthenium deficiency increases the resistivity and reduces the Curie temperature of SrRuO$_3$ and is detrimental to superconductivity in Sr$_2$RuO$_4$.
In extreme cases, SrRuO$_3$ and Sr$_2$RuO$_4$ even show semiconducting behavior at low temperatures \cite{Cao2016,Dabrowski2004}.
Superconductivity in Sr$_2$RuO$_4$ is extremely sensitive to defects, such as nonmagnetic impurities and lattice imperfections and thus it requires high quality samples \cite{Mao1999,Mackenzie1998}.
In order to overcome the volatility of ruthenium oxides, some of us have recently used an adsorption-controlled growth technique to synthesize ruthenate thin films showing superconductivity and unprecedentedly high residual resistivity ratios, defined as the ratio of the resistivities at 300 and 4K \cite{Nair2018-113,Nair2018-214}.
Related results have also been reported by other groups where the growth of Sr$_2$RuO$_4$ films was carried out by means of molecular beam epitaxy \cite{Marshall2017,Uchida2017}.

\begin{figure*}[t]
	\includegraphics[width=7 in]{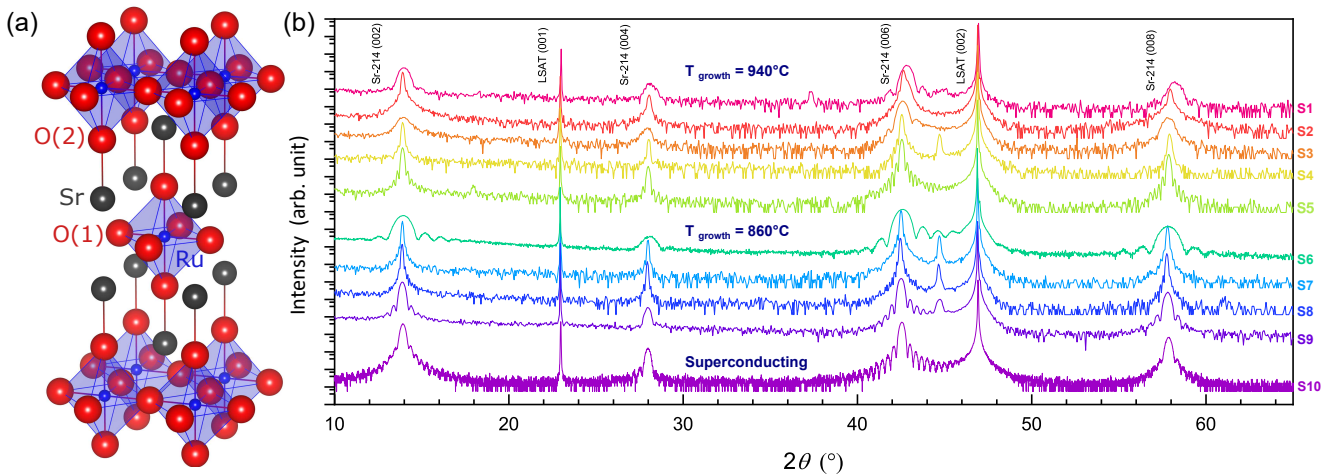}% Here is how to import EPS art
	\caption{\label{fig:structure}(a) The lattice structure of Sr$_2$RuO$_4$ with tetragonal symmetry. (b) XRD $\theta$-2$\theta$ scans of thin films. The sharp peak between Sr-214 (006) and LSAT (002) is the so-called hybrid reflection from combined diffraction of the film and substrate \cite{Dietl2018}. The $y$-axis is logarithmic and the curves are shifted in the $y$-direction for better visibility.}
\end{figure*}

Nevertheless, a method to directly detect the ruthenium vacancies is still lacking due to their atomic dimensions and low concentration.
Until now, the most common approach is based on using the residual resistivity that increases when defects are introduced. 
However, this method can only provide indirect evidence of Ru vacancies, and different factors (such as microstructural defects) can also contribute to the residual resistivity. 
To directly and specifically detect the ruthenium vacancies, a characterization tool that is sensitive to the local crystalline environment is required.

Here we report a combination of spectroscopic methods to directly identify ruthenium vacancies in Sr$_2$RuO$_4$.
Specifically, we utilized X-ray absorption spectroscopy (XAS) and Raman spectroscopy to investigate thin films grown by the adsorption-controlled growth technique using either reactive sputtering or molecular-beam epitaxy (MBE). 
These complementary spectroscopic methods provide information on the modified hybridization between ruthenium and oxygen orbitals and on the local lattice distortions induced by ruthenium deficiencies, respectively.
The spectroscopic signatures of ruthenium vacancies discussed in this work can thus serve as a guiding tool for the growth of ruthenate thin films with controlled stoichiometry, which will benefit both fundamental studies and oxide electronics applications. 

%%%%%%%%%%%%%%%%%%%%%%%% Exp details %%%%%%%%%%%%%%%%%%%%%%%%%%
%%%%%%%%%%%%%%%%%%%%%%%%%%%%%%%%%%%%%%%%%%%%%%%%%%%%%%%%%%%%%%%
\section{Experimental details}
Thin films were grown on (LaAlO$_3$)$_{0.3}$-(SrAl$_{0.5}$Ta$_{0.5}$O$_3$)$_{0.7}$ (LSAT) (001) single-crystalline substrates (CrysTec GmbH) using either a reactive sputtering system developed at the Max Planck Institute for Solid State Research or an oxide MBE system at Cornell University. 
For reactive sputtering, argon and oxygen gas were supplied via a mass flow controller. 
The pressures P$_{O_2}$ and P$_{total}$ were 50 and 100 mTorr, respectively. 
Substrates were glued with a platinum paste to pure nickel blocks and heated with an infrared laser. 
The substrate temperature was monitored using a radiative pyrometer using the emissivity of $\epsilon_{LSAT}$=0.92. 
The structural quality of the films was confirmed by high-resolution X-ray diffraction (XRD) with a Cu K-$\alpha$ source ($\lambda$ $\sim$ 1.5406 \r{A}) and by transmission electron microscopy. 
The growth parameters for the oxide MBE films have been presented elsewhere \cite{Nair2018-214}.
All the samples investigated in this study are listed in table \ref{tab:table1}. 

\begin{table}[h]%The best place to locate the table environment is directly after its first reference in text
	\caption{\label{tab:table1}%
		List of samples. 
		The samples are listed in order of their residual resistivity ratios (see Fig. \ref{fig:transport}). Samples labeled ``metal-insulating" show a resistivity minimum as a function of temperature (Fig. 2). The thicknesses were calculated using Laue fringes in XRD $\theta$-2$\theta$ scans. 
	}
	\begin{ruledtabular}
		\begin{tabular}{ccccc}
			\textrm{Sample}&
			\textrm{Transport}&
			\textrm{T$_{growth}$ (\degc)}&
			\textrm{\textit{d} (nm)}&
			\textrm{Technique}\\
			\colrule
			S1& Insulating & 940& 16 & Sputtering \\
			S2& Metal-insulating&940&  48 & Sputtering  \\
			S3& Metal-insulating &940&  16 & Sputtering  \\
			S4& Metal-insulating &940&  50 & Sputtering  \\
			S5& Metal-insulating &940&  42 & Sputtering  \\
			\colrule
			S6& Metal-insulating &860&  11 & Sputtering  \\
			S7& Metal-insulating &860&  62 & Sputtering  \\
			S8& Metal-insulating &860&  67 & Sputtering  \\
			S9& Metallic &860&  26 & Sputtering \\
			\colrule
			S10& Superconducting &860&  28 & MBE \\
		\end{tabular}
	\end{ruledtabular}
\end{table}

The electric transport measurements were carried out using a Physical Property Measurement System (Quantum Design Co.). To implement the van der Pauw geometry, Ag/Au metallic contacts were deposited with a sputtering on four corners of square shaped samples (5 mm $\times$ 5 mm).
The values of resistivity at room temperature (300 K) are 48255.5, 1126.4, 651.1, 640.9, 457.1, 245.4, 189.3, 171.0, 211.7 $\mu \Omega \cdot$cm in \textit{S1-S9}, respectively.

The Raman spectra were measured with a Jobin-Yvon LabRam HR800 spectrometer (Horiba Co.) combined with a dedicated confocal microscope with the 100$\times$ long working distance objective lens. The short depth of focus allows measurements of films with thicknesses of $\sim$10 nm. The samples were illuminated with a He-Ne laser with wavelength 632.8 nm (red), and the scattered light was collected from the sample surface with a 100$\times$ objective. The experiments were performed in backscattering geometry with (\textit{a},\textit{b})-axis polarized light propagating along the crystallographic \textit{c}-axis, which is denoted as $z(XX)\bar{z}$ in Porto's notation.

The XAS measurements were carried out at the O K-edge at the BL29-BOREAS beamline at the ALBA synchrotron light source (Barcelona, Spain) \cite{Barla2016}. The spectra were measured in total electron yield (TEY) mode under ultrahigh vacuum conditions (1.5 $\times$ 10$^{-10}$ Torr).

For scanning transmission electron microscopy (STEM), we prepared representative cross-sectional electron transparent specimens by employing the standard specimen preparation procedure including mechanical grinding, tripod wedge polishing, and argon ion milling. 
After the specimens were thinned down to $\sim$10 $\mu$m by tripod polishing, argon ion beam milling, for which a precision ion polishing system (PIPS II, Model 695) was used at low temperature, was carried out until reaching electron transparency. 
For all STEM analyses, a probe-aberration-corrected JEOL JEM-ARM200F equipped with a cold field-emission electron source, a probe Cs-corrector (DCOR, CEOS GmbH) and a large solid-angle JEOL Centurio SDD-type energy-dispersive X-ray spectroscopy (EDXS) detector was used. 
The collection angle range for high-angle annular dark-field (HAADF) images was 75-310 mrad. To decrease the noise level, the images were processed with a principal component analysis routine.

%%%%%%%%%%%%%%%%%%%%Result and discussion%%%%%%%%%%%%%%%%%%%%%%
%%%%%%%%%%%%%%%%%%%%%%%%%%%%%%%%%%%%%%%%%%%%%%%%%%%%%%%%%%%%%%%
\section{Results and discussion}
Since ruthenium deficiency in Sr$_2$RuO$_4$ (SRO) thin films is strongly dependent on the growth conditions, the adsorption-controlled growth that employs a large flux of ruthenium is an appropriate way to control the stoichiometry \cite{Nair2018-214}. 
For adsorption-controlled growth of oxides, an oxidation agent is required to evaporate excessive ruthenium as some form of RuO$_x$(g) and to fully oxidize the films.
For this purpose, we used concentrated (distilled) ozone in oxide MBE and an oxygen plasma in reactive sputtering. 
We sputtered SRO thin films with different concentrations of ruthenium defects by tuning two growth parameters, namely the deposition temperature and the ratio between ablations of stoichiometric SRO and ruthenium metal targets.
We kept the oxygen partial pressure of the chamber constant during growths and rapid cooling processes to avoid variations in oxygen content.
In order to obtain the insulating sample \textit{S1} with the highest concentration of ruthenium vacancies, we prepared the sample 940 \degc $ $ using only the SRO target, exploiting the fact that more ruthenium vacancies are generated at higher temperatures.

We investigated the structure of all SRO films using XRD. 
Figure \ref{fig:structure} (b) shows the 00L reflections from LSAT and SRO observed in $\theta$-2$\theta$ scans which demonstrate that the films were oriented along the $c$ axis without parasitic phases.
All curves show Laue fringes, which indicates smooth surfaces and interfaces.
Reflections from different samples show nearly identical peak positions with only minor variations, except for the sample \textit{S1} that has a smaller $c$ lattice parameter due to the excessive ruthenium deficiency. 

\begin{figure}[h]
	\includegraphics[width=3 in]{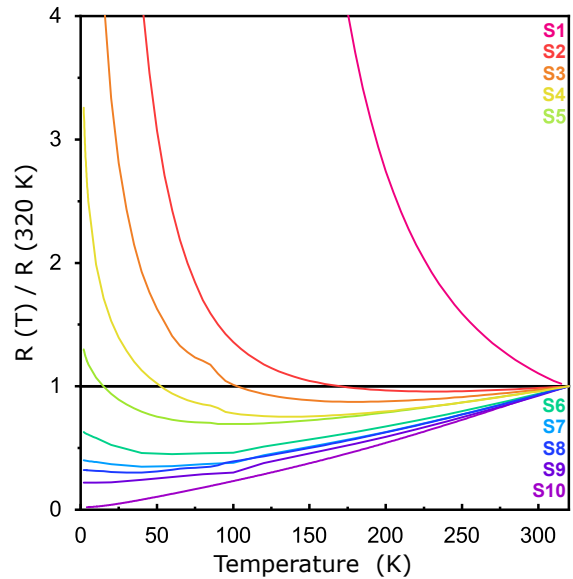}% Here is how to import EPS art
	\caption{\label{fig:transport}Normalized resistance curves. The kink at 80 K in some of the curves is an instrumental artefact.}
\end{figure}

Although the XRD curves only display minor differences, the transport properties vary substantially, ranging from insulating to metallic and finally to superconducting behavior (Fig.\ref{fig:transport}).
The resistance curves were normalized to the resistance at 320 K to highlight the differences in the low-temperature properties.
One sample, \textit{S1}, exhibits a completely insulating behavior that has not been observed before except for exfoliated nano-sized single crystals \cite{Nobukane2017}.
This is consistent with the expectation of high ruthenium deficiency in samples grown with extreme growth parameters. 
The samples grown at 940 \degc$ $ (\textit{S1,2,3,4,5}) have higher normalized resistances than the samples grown at 860 \degc $ $ (\textit{S6,7,8,9}), which suggests that the higher growth temperature favors the formation of defects.  
The MBE-grown sample \textit{S10} has a very low residual resistivity and is superconducting at T$_{c,midpoint}$=0.67 K.

\begin{figure}[h]
	\includegraphics[width=3.45 in]{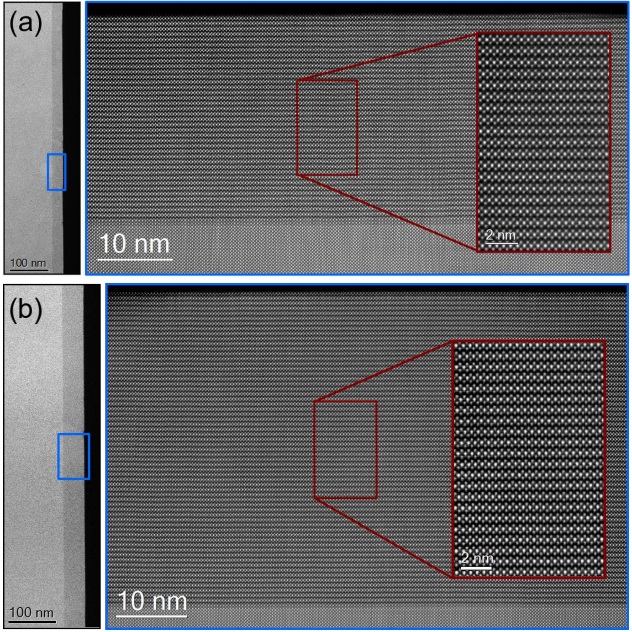}% Here is how to import EPS art
	\caption{\label{fig:TEM}HAADF images for (a) \textit{S9} and (b) \textit{S5} samples. The blue and red squares indicate the magnified regions. The brighter spots are from ruthenium atoms and the other spots are from strontium atoms.}
\end{figure}

We collected STEM-HAADF images to investigate if and how details of the structure could result in different transport properties.
Two representative samples, \textit{S9} and \textit{S5}, that were grown at different temperatures were examined (Fig. \ref{fig:TEM}).
Although the two samples show very different transport properties, the STEM images are strikingly similar. 
Both samples show perfect epitaxial qualities demonstrated by the absence of parasitic phases and structural defects over hundreds of nanometers.
The high magnification HAADF images show the ideal K$_2$NiF$_4$ structure, consistent with the hypothesis that the different transport properties arise from atomic-scale defects. 
The seemingly ideal structures of both fully metallic and metal-insulating samples imply that the insulating behavior does not stem from parasitic phases such as Sr$_3$Ru$_2$O$_7$ or SrRuO$_3$.

\begin{figure}[t]
	\includegraphics[width=3.4 in]{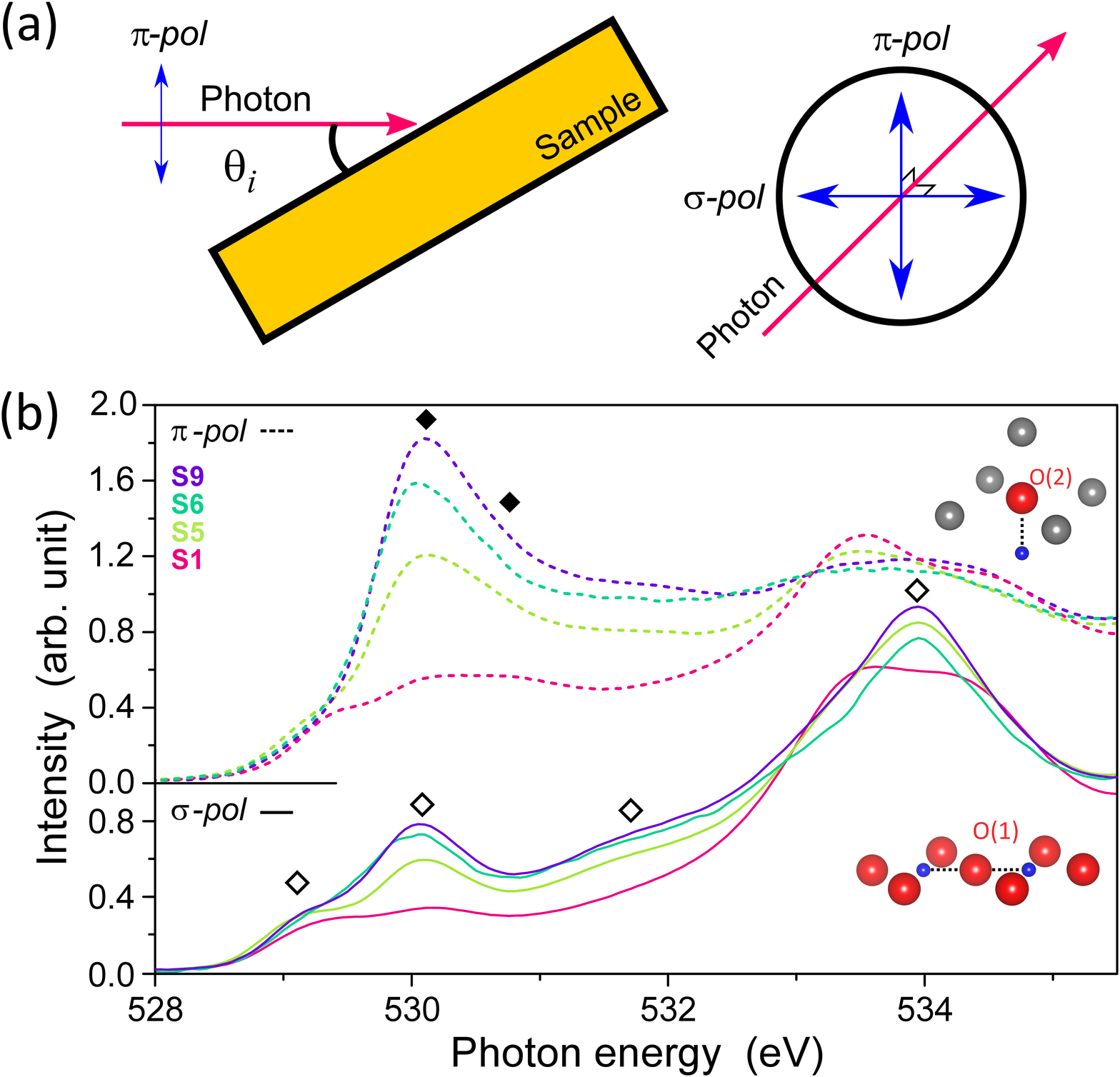}% Here is how to import EPS art
	\caption{\label{fig:XAS}(a) Schematic diagram of the XAS measurement geometry. (b) X-ray absorption near O K-edge spectra. The full and empty diamond shapes indicate peaks from O(2) and O(1), respectively. The insets show the bonding configuration of O(1) and O(2) with Ru. The blue spheres are ruthenium atoms and the grey spheres are strontium atoms consistent with Fig. \ref{fig:structure}(a).}
\end{figure}

Atomic-scale point defects can be studied by core level spectroscopy thanks to its sensitivity to the bonding environment of an atom.
Metal vacancies in metal oxides affect the chemical properties of the materials by reducing the number of metal-oxygen bonds, and raising the oxidation state as the number of oxygen atoms per metal increases.
The higher oxidation state and the reduced number of bonds reduce the hybridization of the metal ion with the surrounding ligands.
In SRO, ruthenium is hybridized with oxygen ions in a [RuO$_6$]$^{4-}$ complex, and this is reflected in O K-edge X-ray absorption spectra as pre-edge peaks \cite{Schmidt1996,Moon2006,Pchelkina2007}.
There are two inequivalent oxygens in SRO: a planar oxygen labeled as O(1), and an apical oxygen labeled as O(2) (Fig. \ref{fig:structure}(a) and inset to Fig. \ref{fig:XAS}(b)).
XAS with linearly polarized light can resolve the difference between O(1) and O(2) thanks to the angle-dependence of the signal, as X-ray absorption occurs selectively in orbitals that lie parallel to the electric-field vector of the incoming x-rays.

We carried out linearly polarized XAS measurements near the O K-edge in TEY mode on four of our films at room temperature (Fig. \ref{fig:XAS}).
The $c$ axis of the samples is perpendicular to the surface, so that the spectra are more sensitive to the planar O(1) at normal incidence, $\theta_i$ = 90 $ $\dego, and  pick up substantial contributions from the O(2) as $\theta_i$ approaches grazing incidence, towards $\theta_i$ = 0 $ $\dego$ $ (Figure. \ref{fig:XAS}(a)).
However, instead of changing the incident angle, we fixed it at 30 $ $\dego $ $ and controlled the polarization of the incident x-rays.
In this configuration, the vertical polarization ($\sigma$-pol) is always parallel to the Ru-O(1) bond, and the horizontal polarization ($\pi$-pol) picks up contributions from both Ru-O(1) and Ru-O(2) bonds. Given the 30 $ $\dego $ $ angle from Ru-O(2), the signal from O(2) is dominant.
The advantage of this configuration is that we can minimize variations of the footprint area of the X-ray beam, by keeping the incident angle fixed during the measurements.

Four samples with different transport properties were chosen for the measurements. 
The metallic sample \textit{S9} shows spectra consistent with the previously reported spectra of single crystals, thereby confirming a good sample quality \cite{Moon2006, Schmidt1996}. On the other hand, the less conducting samples \textit{S6} and \textit{S5} exhibit spectra with suppressed pre-peak spectral weight.
Especially the spectral weight between 529 eV and 533 eV is progressively suppressed in measurements with $\pi$-polarization.
On the other hand, the peak at 529 eV and 534 eV in the $\sigma$-polarized spectra shows a smaller suppression.
The spectra from the insulating sample \textit{S1} reveal a somewhat different shape from the other three samples due to the high concentration of ruthenium vacancies and its insulating behavior at room temperature, and also show greatly suppressed spectral weight at the pre-edge.

The change in spectral weight in the pre-edge structure can be interpreted as a signature of the ruthenium vacancies due to the above-mentioned reduced number of bonds and diminished hybridization.
In particular, the stronger suppression of the out-of-plane component is remarkable and can be explained by a different number of ruthenium atoms in proximity to O(1) and O(2).
As depicted in the inset of Fig. \ref{fig:XAS}(b), the O(1) has two bonds with neighboring ruthenium atoms in the $xy$ plane, whereas O(2) has only one bond with Ru in the $z$-direction. The apical oxygen is therefore more sensitive to the presence of ruthenium vacancies.

Now we shift our attention to the local lattice distortion produced by the ruthenium vacancies. 
The absence of a ruthenium atom not only generates a void, but also distorts the lattice around the vacancy site, which results in lower local lattice symmetry.
Raman spectroscopy is a suitable tool to study the change in local symmetries in crystals, because the Raman-activity of a phonon mode can be determined by a group-theoretical analysis of the lattice.
SRO has a simple lattice structure with tetragonal symmetry, and the Ru and O(1) atoms are located in centrosymmetric positions.
Due to the high symmetry, only four Raman modes (2$A_{1g}$+2$E_g$) are expected from group theory \cite{Iliev2005}.

The recent development of confocal Raman spectroscopy, based on an optical microscope with motorized objective lens and a confocal hole to reject substrate contributions, has provided new information on thin films down to thicknesses of a few nanometers, including the strain dependence of charge ordering phenomena, thickness dependent lattice structures, and oxygen vacancies \cite{Hepting2014,Kim2017}.
We used a confocal micro-Raman setup to examine the local structural change in ruthenium deficient SRO thin films, where the signal from the thin film can be extracted using a depth-resolved measurement \cite{Hepting2014}.
The thin film shape restricted the measurement to the $z(XX)\bar{z}$ geometry, in which the propagation of light is parallel to the surface normal, and the polarization of the light is parallel to the $a$ axis of the lattice. 
In this geometry we could study phonon modes with $A_{1g}+ B_{1g}$ symmetry, including the two $A_{1g}$ phonon modes at 200 \wn$ $ and 545 \wn$ $ that are related to the vibration of Sr and O(2) in the $z$ direction \cite{Iliev2005,Sakita2001}.

\begin{figure}[h]
	\includegraphics[width=3.3 in]{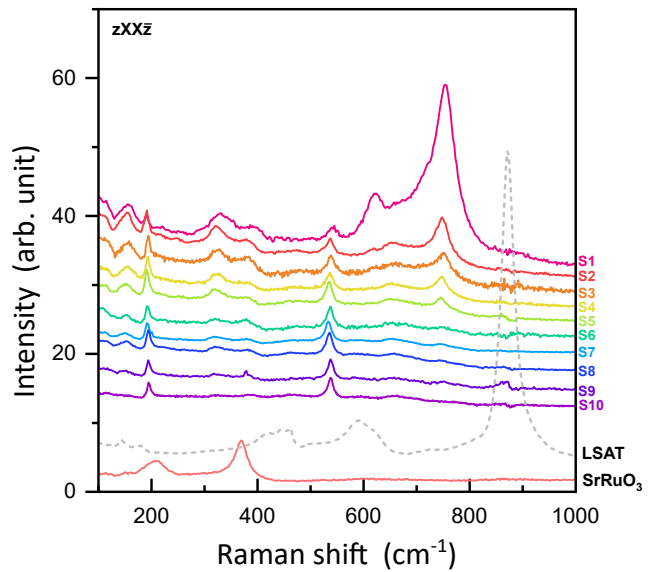}% Here is how to import EPS art
	\caption{\label{fig:RamanAll}Polarized Raman spectra acquired at room temperature. The gray dashed curve is the LSAT substrate signal that was subtracted out in the analysis.}
\end{figure}

\begin{figure}[h]
	\includegraphics[width=3.3 in]{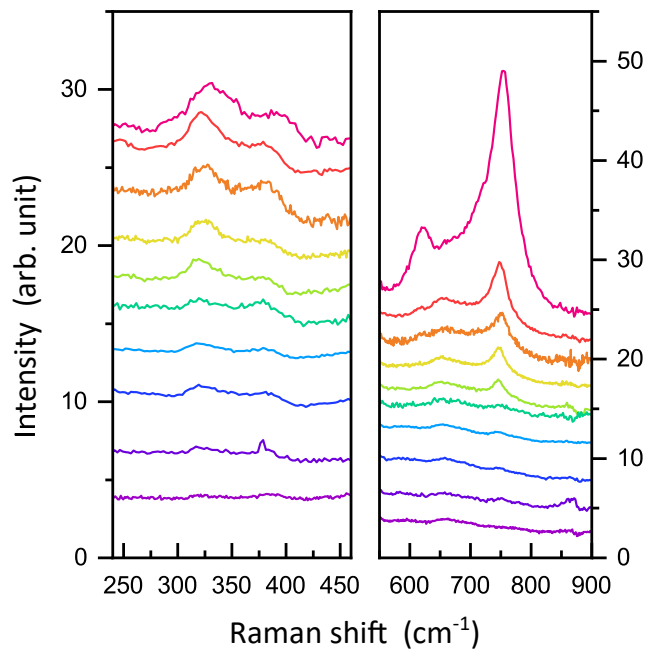}% Here is how to import EPS art
	\caption{\label{fig:RamanZoom} Detailed view of the Raman features attributed to Ru vacancies in Fig. \ref{fig:RamanAll}.}
\end{figure}

In Figure \ref{fig:RamanAll}, we present substrate-subtracted Raman spectra acquired at room temperature for all investigated samples. The spectra were normalized to the peak at 545 \wn$ $, and shifted along the $y$-axis for the sake of readability.
The metallic and superconducting samples, \textit{S9} and \textit{S10}, show two sharp peaks from $A_{1g}$ phonon modes, as observed in single crystals \cite{Iliev2005,Sakita2001}.
In less metallic samples, additional peaks appear in two energy windows: between 300 and 400 \wn$ $, and between 600 and 800 \wn$ $.
The spectra from the substrate and SrRuO$_3$ are plotted in Fig. 5 to demonstrate that the additional features are neither originating from the substrate nor from impurity phases.
The spectral weight of the additional features grows as the samples become more insulating, thereby revealing a strong correlation between the Raman spectra and the transport properties. 
As a consequence we can safely assign the new peaks to Raman-active phonon modes originating from the locally reduced symmetry induced by ruthenium vacancies. 

In order to investigate the nature of the new features, magnified spectra are displayed in Fig. \ref{fig:RamanZoom}.
The energies and spectral shapes of the new features suggest that peaks at higher energy (\textit{E}), $600$ $ $ \wn$ < E < 800$$ $ \wn$ $, are overtones of the phonon modes at lower energy, $300$ $ $ \wn$ < E < 400$$ $ \wn$ $. 
Specifically, the former peaks are seen at exactly twice the energies of the latter peaks, as expected for second-order overtone modes in the harmonic approximation.
Moreover, the intensities of all new peaks grow proportionately in less conducting samples, which points to a common underlying origin.
Ruthenium vacancies thus generate at least two phonon modes in the range $300$ $ $ \wn$ < E < 400$$ $ \wn$ $, and their overtones are visible in the spectra as well.
A possible origin of the additional peaks is the mixing of Raman and infrared (IR) active modes due to the reduced local symmetry, in which the IR-active modes can appear in the Raman spectra.
Indeed, two IR active modes were observed between 300 and 400 \wn $ $  by inelastic neutron scattering \cite{Braden2007}. 

\begin{figure}[h]
	\includegraphics[width=3.2 in]{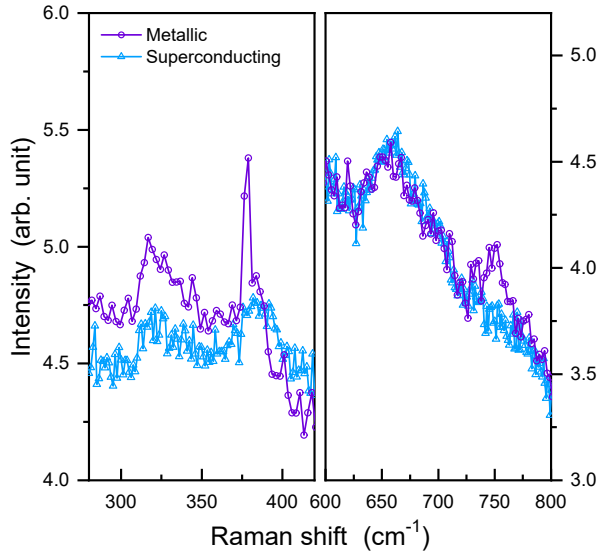}% Here is how to import EPS art
	\caption{\label{fig:RamanCompare} Detailed view of the specral ranges exhibiting signatures of Ru vacancies in the Raman spectra of the metallic sample \textit{S9} and the superconducting sample \textit{S10}.}
\end{figure}

In addition, when looking closely at the samples \textit{S9} and \textit{S10} showing metallic transport properties, one can observe faint but visible peaks at the same energies, which indicate a small concentration of Ru vacancies (Fig. \ref{fig:RamanCompare}).
This suggests that even the best Sr$_2$RuO$_4$ thin-film samples incorporate a small number of point defects during the high temperature growth process in high vacuum. 
We also note small but distinct differences in the spectra of the metallic and superconducting samples: the peak at 375 \wn$ $ and its overtone at 750 \wn$ $ are visible in the spectra of sample \textit{S9}, but not in those of \textit{S10}.

%%%%%%%%%%%%%%%%%%%%%%%% paragraph 7 %%%%%%%%%%%%%%%%%%%%%%%%%%

\section{Conclusion}
In conclusion, our study has provided two spectroscopic signatures of ruthenium vacancies in SRO thin films that are not detectable with other state-of-art techniques, such as STEM and high resolution XRD.
The ruthenium deficiency of the films was tuned by varying the growth temperature as well as growth technique.
The resulting films exhibit widely different transport properties, ranging from insulating all the way to superconducting behavior.
In ruthenium deficient SRO, the O K-edge XAS exhibits a considerable suppression of the spectral weight in the pre-peak structures stemming from Ru-O hybridization.
We also used confocal micro-Raman spectroscopy to characterize additional phonon modes arising from point defects, following related observations on bulk materials \cite{Ammundsen1999,Limmer1998} and copper oxide thin films \cite{Kim2017}.
Our spectroscopic results call for further quantitative analyses on ruthenium vacancies in SRO thin films using techniques like the Rutherford backscattering measurement, which can calibrate the correlation between the spectroscopic features and the density of ruthenium vacancies. 
We expect that our results will be helpful in guiding the preparation of ruthenate thin films for fundamental physics and applications in oxide electronics.

\begin{acknowledgments}
	We thank R. Usiskin and E. Heifets for fruitful discussion, B. Stuhlhofer, P. Specht, S. Schmid, B. Lemke and M. Schultz for technical support, and U. Salzberger for TEM specimen preparation. We used VESTA software to generate the atomic models used in this article.
	The XAS experiments were performed at beamline BL29 at the ALBA synchrotron with support from the beamline staffs.
	
	The project was supported by the European Research Council under Advanced Grant No. 669550 (Com4Com).
	
	P.A.vA and Y.E.S acknowledge support from the European Union's Horizon 2020 research and innovation programme under grant agreement No. 823717 - ESTEEM3. 
	The work at Cornell was supported by the National Science Foundation (Platform for Accelerated Realization, Analysis and Discovery of Interface Materials (PARADIM)) under Cooperative Agreement No. DMR-1539918, the W.M. Keck Foundation, and the Gordon and Betty Moore Foundation's EPiQS Initiative through Grant No. GBMF3850. N.J.S. acknowledges support from the National Science Foundation Graduate Research Fellowship Program under Grant No. DGE-1650441.
	
\end{acknowledgments}

\bibliography{apssamp}% Produces the bibliography via BibTeX.

\providecommand{\noopsort}[1]{}\providecommand{\singleletter}[1]{#1}%
\begin{thebibliography}{38}%
\makeatletter
\providecommand \@ifxundefined [1]{%
 \@ifx{#1\undefined}
}%
\providecommand \@ifnum [1]{%
 \ifnum #1\expandafter \@firstoftwo
 \else \expandafter \@secondoftwo
 \fi
}%
\providecommand \@ifx [1]{%
 \ifx #1\expandafter \@firstoftwo
 \else \expandafter \@secondoftwo
 \fi
}%
\providecommand \natexlab [1]{#1}%
\providecommand \enquote  [1]{``#1''}%
\providecommand \bibnamefont  [1]{#1}%
\providecommand \bibfnamefont [1]{#1}%
\providecommand \citenamefont [1]{#1}%
\providecommand \href@noop [0]{\@secondoftwo}%
\providecommand \href [0]{\begingroup \@sanitize@url \@href}%
\providecommand \@href[1]{\@@startlink{#1}\@@href}%
\providecommand \@@href[1]{\endgroup#1\@@endlink}%
\providecommand \@sanitize@url [0]{\catcode `\\12\catcode `\$12\catcode
  `\&12\catcode `\#12\catcode `\^12\catcode `\_12\catcode `\%12\relax}%
\providecommand \@@startlink[1]{}%
\providecommand \@@endlink[0]{}%
\providecommand \url  [0]{\begingroup\@sanitize@url \@url }%
\providecommand \@url [1]{\endgroup\@href {#1}{\urlprefix }}%
\providecommand \urlprefix  [0]{URL }%
\providecommand \Eprint [0]{\href }%
\providecommand \doibase [0]{https://doi.org/}%
\providecommand \selectlanguage [0]{\@gobble}%
\providecommand \bibinfo  [0]{\@secondoftwo}%
\providecommand \bibfield  [0]{\@secondoftwo}%
\providecommand \translation [1]{[#1]}%
\providecommand \BibitemOpen [0]{}%
\providecommand \bibitemStop [0]{}%
\providecommand \bibitemNoStop [0]{.\EOS\space}%
\providecommand \EOS [0]{\spacefactor3000\relax}%
\providecommand \BibitemShut  [1]{\csname bibitem#1\endcsname}%
\let\auto@bib@innerbib\@empty
%</preamble>
\bibitem [{\citenamefont {Mackenzie}\ \emph {et~al.}(1996)\citenamefont
  {Mackenzie}, \citenamefont {Julian}, \citenamefont {Diver}, \citenamefont
  {McMullan}, \citenamefont {Ray}, \citenamefont {Lonzarich}, \citenamefont
  {Maeno}, \citenamefont {Nishizaki},\ and\ \citenamefont
  {Fujita}}]{Mackenzie1996}%
  \BibitemOpen
  \bibfield  {author} {\bibinfo {author} {\bibfnamefont {A.~P.}\ \bibnamefont
  {Mackenzie}}, \bibinfo {author} {\bibfnamefont {S.~R.}\ \bibnamefont
  {Julian}}, \bibinfo {author} {\bibfnamefont {A.~J.}\ \bibnamefont {Diver}},
  \bibinfo {author} {\bibfnamefont {G.~J.}\ \bibnamefont {McMullan}}, \bibinfo
  {author} {\bibfnamefont {M.~P.}\ \bibnamefont {Ray}}, \bibinfo {author}
  {\bibfnamefont {G.~G.}\ \bibnamefont {Lonzarich}}, \bibinfo {author}
  {\bibfnamefont {Y.}~\bibnamefont {Maeno}}, \bibinfo {author} {\bibfnamefont
  {S.}~\bibnamefont {Nishizaki}},\ and\ \bibinfo {author} {\bibfnamefont
  {T.}~\bibnamefont {Fujita}},\ }\href@noop {} {\bibfield  {journal} {\bibinfo
  {journal} {Phys. Rev. Lett.}\ }\textbf {\bibinfo {volume} {76}},\ \bibinfo
  {pages} {3786} (\bibinfo {year} {1996})}\BibitemShut {NoStop}%
\bibitem [{\citenamefont {Mackenzie}\ \emph {et~al.}(2017)\citenamefont
  {Mackenzie}, \citenamefont {Scaffidi}, \citenamefont {Hicks},\ and\
  \citenamefont {Maeno}}]{Mackenzie2017}%
  \BibitemOpen
  \bibfield  {author} {\bibinfo {author} {\bibfnamefont {A.~P.}\ \bibnamefont
  {Mackenzie}}, \bibinfo {author} {\bibfnamefont {T.}~\bibnamefont {Scaffidi}},
  \bibinfo {author} {\bibfnamefont {C.~W.}\ \bibnamefont {Hicks}},\ and\
  \bibinfo {author} {\bibfnamefont {Y.}~\bibnamefont {Maeno}},\ }\href@noop {}
  {\bibfield  {journal} {\bibinfo  {journal} {npj Quantum Materials}\ }\textbf
  {\bibinfo {volume} {2}},\ \bibinfo {pages} {40} (\bibinfo {year}
  {2017})}\BibitemShut {NoStop}%
\bibitem [{\citenamefont {Pustogow}\ \emph {et~al.}()\citenamefont {Pustogow},
  \citenamefont {Luo}, \citenamefont {Chronister}, \citenamefont {Su},
  \citenamefont {Sokolov}, \citenamefont {Jerzembeck}, \citenamefont
  {Mackenzie}, \citenamefont {Hicks}, \citenamefont {Kikugawa}, \citenamefont
  {Raghu}, \citenamefont {Bauer},\ and\ \citenamefont {Brown}}]{Pustogow2019}%
  \BibitemOpen
  \bibfield  {author} {\bibinfo {author} {\bibfnamefont {A.}~\bibnamefont
  {Pustogow}}, \bibinfo {author} {\bibfnamefont {Y.}~\bibnamefont {Luo}},
  \bibinfo {author} {\bibfnamefont {A.}~\bibnamefont {Chronister}}, \bibinfo
  {author} {\bibfnamefont {Y.-S.}\ \bibnamefont {Su}}, \bibinfo {author}
  {\bibfnamefont {D.~A.}\ \bibnamefont {Sokolov}}, \bibinfo {author}
  {\bibfnamefont {F.}~\bibnamefont {Jerzembeck}}, \bibinfo {author}
  {\bibfnamefont {A.~P.}\ \bibnamefont {Mackenzie}}, \bibinfo {author}
  {\bibfnamefont {C.~W.}\ \bibnamefont {Hicks}}, \bibinfo {author}
  {\bibfnamefont {N.}~\bibnamefont {Kikugawa}}, \bibinfo {author}
  {\bibfnamefont {S.}~\bibnamefont {Raghu}}, \bibinfo {author} {\bibfnamefont
  {E.~D.}\ \bibnamefont {Bauer}},\ and\ \bibinfo {author} {\bibfnamefont
  {S.~E.}\ \bibnamefont {Brown}},\ }\href@noop {} {\bibinfo  {journal}
  {arxiv:1904.00047 (2019)}\ }\BibitemShut {NoStop}%
\bibitem [{\citenamefont {Steppke}\ \emph {et~al.}(2017)\citenamefont
  {Steppke}, \citenamefont {Zhao}, \citenamefont {Barber}, \citenamefont
  {Scaffidi}, \citenamefont {Jerzembeck}, \citenamefont {Rosner}, \citenamefont
  {Gibbs}, \citenamefont {Maeno}, \citenamefont {Simon}, \citenamefont
  {Mackenzie},\ and\ \citenamefont {Hicks}}]{Steppke2017}%
  \BibitemOpen
\bibfield  {journal} {  }\bibfield  {author} {\bibinfo {author} {\bibfnamefont
  {A.}~\bibnamefont {Steppke}}, \bibinfo {author} {\bibfnamefont
  {L.}~\bibnamefont {Zhao}}, \bibinfo {author} {\bibfnamefont {M.~E.}\
  \bibnamefont {Barber}}, \bibinfo {author} {\bibfnamefont {T.}~\bibnamefont
  {Scaffidi}}, \bibinfo {author} {\bibfnamefont {F.}~\bibnamefont
  {Jerzembeck}}, \bibinfo {author} {\bibfnamefont {H.}~\bibnamefont {Rosner}},
  \bibinfo {author} {\bibfnamefont {A.~S.}\ \bibnamefont {Gibbs}}, \bibinfo
  {author} {\bibfnamefont {Y.}~\bibnamefont {Maeno}}, \bibinfo {author}
  {\bibfnamefont {S.~H.}\ \bibnamefont {Simon}}, \bibinfo {author}
  {\bibfnamefont {A.~P.}\ \bibnamefont {Mackenzie}},\ and\ \bibinfo {author}
  {\bibfnamefont {C.~W.}\ \bibnamefont {Hicks}},\ }\href@noop {} {\bibfield
  {journal} {\bibinfo  {journal} {Science}\ }\textbf {\bibinfo {volume}
  {355}},\ \bibinfo {pages} {148} (\bibinfo {year} {2017})}\BibitemShut
  {NoStop}%
\bibitem [{\citenamefont {Barber}\ \emph {et~al.}(2018)\citenamefont {Barber},
  \citenamefont {Gibbs}, \citenamefont {Maeno}, \citenamefont {Mackenzie},\
  and\ \citenamefont {Hicks}}]{Barber2018}%
  \BibitemOpen
  \bibfield  {author} {\bibinfo {author} {\bibfnamefont {M.~E.}\ \bibnamefont
  {Barber}}, \bibinfo {author} {\bibfnamefont {A.~S.}\ \bibnamefont {Gibbs}},
  \bibinfo {author} {\bibfnamefont {Y.}~\bibnamefont {Maeno}}, \bibinfo
  {author} {\bibfnamefont {A.~P.}\ \bibnamefont {Mackenzie}},\ and\ \bibinfo
  {author} {\bibfnamefont {C.~W.}\ \bibnamefont {Hicks}},\ }\href
  {https://doi.org/10.1103/PhysRevLett.120.076602} {\bibfield  {journal}
  {\bibinfo  {journal} {Phys. Rev. Lett.}\ }\textbf {\bibinfo {volume} {120}},\
  \bibinfo {pages} {076602} (\bibinfo {year} {2018})}\BibitemShut {NoStop}%
\bibitem [{\citenamefont {Burganov}\ \emph {et~al.}(2016)\citenamefont
  {Burganov}, \citenamefont {Adamo}, \citenamefont {Mulder}, \citenamefont
  {Uchida}, \citenamefont {King}, \citenamefont {Harter}, \citenamefont {Shai},
  \citenamefont {Gibbs}, \citenamefont {Mackenzie}, \citenamefont {Uecker},
  \citenamefont {Bruetzam}, \citenamefont {Beasley}, \citenamefont {Fennie},
  \citenamefont {Schlom},\ and\ \citenamefont {Shen}}]{Burganov2016}%
  \BibitemOpen
  \bibfield  {author} {\bibinfo {author} {\bibfnamefont {B.}~\bibnamefont
  {Burganov}}, \bibinfo {author} {\bibfnamefont {C.}~\bibnamefont {Adamo}},
  \bibinfo {author} {\bibfnamefont {A.}~\bibnamefont {Mulder}}, \bibinfo
  {author} {\bibfnamefont {M.}~\bibnamefont {Uchida}}, \bibinfo {author}
  {\bibfnamefont {P.~D.~C.}\ \bibnamefont {King}}, \bibinfo {author}
  {\bibfnamefont {J.~W.}\ \bibnamefont {Harter}}, \bibinfo {author}
  {\bibfnamefont {D.~E.}\ \bibnamefont {Shai}}, \bibinfo {author}
  {\bibfnamefont {A.~S.}\ \bibnamefont {Gibbs}}, \bibinfo {author}
  {\bibfnamefont {A.~P.}\ \bibnamefont {Mackenzie}}, \bibinfo {author}
  {\bibfnamefont {R.}~\bibnamefont {Uecker}}, \bibinfo {author} {\bibfnamefont
  {M.}~\bibnamefont {Bruetzam}}, \bibinfo {author} {\bibfnamefont {M.~R.}\
  \bibnamefont {Beasley}}, \bibinfo {author} {\bibfnamefont {C.~J.}\
  \bibnamefont {Fennie}}, \bibinfo {author} {\bibfnamefont {D.~G.}\
  \bibnamefont {Schlom}},\ and\ \bibinfo {author} {\bibfnamefont {K.~M.}\
  \bibnamefont {Shen}},\ }\href
  {https://doi.org/10.1103/PhysRevLett.116.197003} {\bibfield  {journal}
  {\bibinfo  {journal} {Phys. Rev. Lett.}\ }\textbf {\bibinfo {volume} {116}},\
  \bibinfo {pages} {197003} (\bibinfo {year} {2016})}\BibitemShut {NoStop}%
\bibitem [{\citenamefont {Hsu}\ \emph {et~al.}(2016)\citenamefont {Hsu},
  \citenamefont {Cho}, \citenamefont {Rebola}, \citenamefont {Burganov},
  \citenamefont {Adamo}, \citenamefont {Shen}, \citenamefont {Schlom},
  \citenamefont {Fennie},\ and\ \citenamefont {Kim}}]{Hsu2016}%
  \BibitemOpen
  \bibfield  {author} {\bibinfo {author} {\bibfnamefont {Y.-T.}\ \bibnamefont
  {Hsu}}, \bibinfo {author} {\bibfnamefont {W.}~\bibnamefont {Cho}}, \bibinfo
  {author} {\bibfnamefont {A.~F.}\ \bibnamefont {Rebola}}, \bibinfo {author}
  {\bibfnamefont {B.}~\bibnamefont {Burganov}}, \bibinfo {author}
  {\bibfnamefont {C.}~\bibnamefont {Adamo}}, \bibinfo {author} {\bibfnamefont
  {K.~M.}\ \bibnamefont {Shen}}, \bibinfo {author} {\bibfnamefont {D.~G.}\
  \bibnamefont {Schlom}}, \bibinfo {author} {\bibfnamefont {C.~J.}\
  \bibnamefont {Fennie}},\ and\ \bibinfo {author} {\bibfnamefont {E.-A.}\
  \bibnamefont {Kim}},\ }\href {https://doi.org/10.1103/PhysRevB.94.045118}
  {\bibfield  {journal} {\bibinfo  {journal} {Phys. Rev. B}\ }\textbf {\bibinfo
  {volume} {94}},\ \bibinfo {pages} {045118} (\bibinfo {year}
  {2016})}\BibitemShut {NoStop}%
\bibitem [{\citenamefont {Koster}\ \emph {et~al.}(2012)\citenamefont {Koster},
  \citenamefont {Klein}, \citenamefont {Siemons}, \citenamefont {Rijnders},
  \citenamefont {Dodge}, \citenamefont {Eom}, \citenamefont {Blank},\ and\
  \citenamefont {Beasley}}]{Koster2012}%
  \BibitemOpen
  \bibfield  {author} {\bibinfo {author} {\bibfnamefont {G.}~\bibnamefont
  {Koster}}, \bibinfo {author} {\bibfnamefont {L.}~\bibnamefont {Klein}},
  \bibinfo {author} {\bibfnamefont {W.}~\bibnamefont {Siemons}}, \bibinfo
  {author} {\bibfnamefont {G.}~\bibnamefont {Rijnders}}, \bibinfo {author}
  {\bibfnamefont {J.~S.}\ \bibnamefont {Dodge}}, \bibinfo {author}
  {\bibfnamefont {C.-B.}\ \bibnamefont {Eom}}, \bibinfo {author} {\bibfnamefont
  {D.~H.~A.}\ \bibnamefont {Blank}},\ and\ \bibinfo {author} {\bibfnamefont
  {M.~R.}\ \bibnamefont {Beasley}},\ }\href
  {https://doi.org/10.1103/RevModPhys.84.253} {\bibfield  {journal} {\bibinfo
  {journal} {Rev. Mod. Phys.}\ }\textbf {\bibinfo {volume} {84}},\ \bibinfo
  {pages} {253} (\bibinfo {year} {2012})}\BibitemShut {NoStop}%
\bibitem [{\citenamefont {Garcia}\ and\ \citenamefont
  {Bibes}(2014)}]{Garcia2014}%
  \BibitemOpen
  \bibfield  {author} {\bibinfo {author} {\bibfnamefont {V.}~\bibnamefont
  {Garcia}}\ and\ \bibinfo {author} {\bibfnamefont {M.}~\bibnamefont {Bibes}},\
  }\href@noop {} {\bibfield  {journal} {\bibinfo  {journal} {Nat. Commun.}\
  }\textbf {\bibinfo {volume} {5}},\ \bibinfo {pages} {4289} (\bibinfo {year}
  {2014})}\BibitemShut {NoStop}%
\bibitem [{\citenamefont {Takahashi}\ and\ \citenamefont
  {Lippmaa}(2017)}]{Takahasi2017}%
  \BibitemOpen
  \bibfield  {author} {\bibinfo {author} {\bibfnamefont {R.}~\bibnamefont
  {Takahashi}}\ and\ \bibinfo {author} {\bibfnamefont {M.}~\bibnamefont
  {Lippmaa}},\ }\href {https://doi.org/10.1021/acsami.7b03577} {\bibfield
  {journal} {\bibinfo  {journal} {ACS Appl. Mater. Interfaces}\ }\textbf
  {\bibinfo {volume} {9}},\ \bibinfo {pages} {21314} (\bibinfo {year}
  {2017})}\BibitemShut {NoStop}%
\bibitem [{\citenamefont {Anwar}\ \emph {et~al.}(2016)\citenamefont {Anwar},
  \citenamefont {Lee}, \citenamefont {Ishiguro}, \citenamefont {Sugimoto},
  \citenamefont {Tano}, \citenamefont {Kang}, \citenamefont {Shin},
  \citenamefont {Yonezawa}, \citenamefont {Manske}, \citenamefont {Takayanagi},
  \citenamefont {Noh},\ and\ \citenamefont {Maeno}}]{Anwar2016}%
  \BibitemOpen
  \bibfield  {author} {\bibinfo {author} {\bibfnamefont {M.~S.}\ \bibnamefont
  {Anwar}}, \bibinfo {author} {\bibfnamefont {S.~R.}\ \bibnamefont {Lee}},
  \bibinfo {author} {\bibfnamefont {R.}~\bibnamefont {Ishiguro}}, \bibinfo
  {author} {\bibfnamefont {Y.}~\bibnamefont {Sugimoto}}, \bibinfo {author}
  {\bibfnamefont {Y.}~\bibnamefont {Tano}}, \bibinfo {author} {\bibfnamefont
  {S.~J.}\ \bibnamefont {Kang}}, \bibinfo {author} {\bibfnamefont {Y.~J.}\
  \bibnamefont {Shin}}, \bibinfo {author} {\bibfnamefont {S.}~\bibnamefont
  {Yonezawa}}, \bibinfo {author} {\bibfnamefont {D.}~\bibnamefont {Manske}},
  \bibinfo {author} {\bibfnamefont {H.}~\bibnamefont {Takayanagi}}, \bibinfo
  {author} {\bibfnamefont {T.~W.}\ \bibnamefont {Noh}},\ and\ \bibinfo {author}
  {\bibfnamefont {Y.}~\bibnamefont {Maeno}},\ }\href@noop {} {\bibfield
  {journal} {\bibinfo  {journal} {Nat. Commun.}\ }\textbf {\bibinfo {volume}
  {7}},\ \bibinfo {pages} {13220} (\bibinfo {year} {2016})}\BibitemShut
  {NoStop}%
\bibitem [{\citenamefont {Chung}\ \emph {et~al.}(2018)\citenamefont {Chung},
  \citenamefont {Kim}, \citenamefont {Lee},\ and\ \citenamefont
  {Tserkovnyak}}]{Chung2018}%
  \BibitemOpen
  \bibfield  {author} {\bibinfo {author} {\bibfnamefont {S.~B.}\ \bibnamefont
  {Chung}}, \bibinfo {author} {\bibfnamefont {S.~K.}\ \bibnamefont {Kim}},
  \bibinfo {author} {\bibfnamefont {K.~H.}\ \bibnamefont {Lee}},\ and\ \bibinfo
  {author} {\bibfnamefont {Y.}~\bibnamefont {Tserkovnyak}},\ }\href
  {https://doi.org/10.1103/PhysRevLett.121.167001} {\bibfield  {journal}
  {\bibinfo  {journal} {Phys. Rev. Lett.}\ }\textbf {\bibinfo {volume} {121}},\
  \bibinfo {pages} {167001} (\bibinfo {year} {2018})}\BibitemShut {NoStop}%
\bibitem [{\citenamefont {Mufazalova}\ \emph {et~al.}(2018)\citenamefont
  {Mufazalova}, \citenamefont {Belozerov},\ and\ \citenamefont
  {Streltsov}}]{Mufazalova2018}%
  \BibitemOpen
  \bibfield  {author} {\bibinfo {author} {\bibfnamefont {A.~O.}\ \bibnamefont
  {Mufazalova}}, \bibinfo {author} {\bibfnamefont {A.~S.}\ \bibnamefont
  {Belozerov}},\ and\ \bibinfo {author} {\bibfnamefont {S.~V.}\ \bibnamefont
  {Streltsov}},\ }\href@noop {} {\bibfield  {journal} {\bibinfo  {journal}
  {Phys. Rev. B}\ }\textbf {\bibinfo {volume} {98}},\ \bibinfo {pages} {134441}
  (\bibinfo {year} {2018})}\BibitemShut {NoStop}%
\bibitem [{\citenamefont {Nair}\ \emph
  {et~al.}(2018{\natexlab{a}})\citenamefont {Nair}, \citenamefont {Liu},
  \citenamefont {Ruf}, \citenamefont {Schreiber}, \citenamefont {Shang},
  \citenamefont {Baek}, \citenamefont {Goodge}, \citenamefont {Kourkoutis},
  \citenamefont {Liu}, \citenamefont {Shen},\ and\ \citenamefont
  {Schlom}}]{Nair2018-113}%
  \BibitemOpen
  \bibfield  {author} {\bibinfo {author} {\bibfnamefont {H.~P.}\ \bibnamefont
  {Nair}}, \bibinfo {author} {\bibfnamefont {Y.}~\bibnamefont {Liu}}, \bibinfo
  {author} {\bibfnamefont {J.~P.}\ \bibnamefont {Ruf}}, \bibinfo {author}
  {\bibfnamefont {N.~J.}\ \bibnamefont {Schreiber}}, \bibinfo {author}
  {\bibfnamefont {S.-L.}\ \bibnamefont {Shang}}, \bibinfo {author}
  {\bibfnamefont {D.~J.}\ \bibnamefont {Baek}}, \bibinfo {author}
  {\bibfnamefont {B.~H.}\ \bibnamefont {Goodge}}, \bibinfo {author}
  {\bibfnamefont {L.~F.}\ \bibnamefont {Kourkoutis}}, \bibinfo {author}
  {\bibfnamefont {Z.-K.}\ \bibnamefont {Liu}}, \bibinfo {author} {\bibfnamefont
  {K.~M.}\ \bibnamefont {Shen}},\ and\ \bibinfo {author} {\bibfnamefont
  {D.~G.}\ \bibnamefont {Schlom}},\ }\href {https://doi.org/10.1063/1.5023477}
  {\bibfield  {journal} {\bibinfo  {journal} {APL Materials}\ }\textbf
  {\bibinfo {volume} {6}},\ \bibinfo {pages} {046101} (\bibinfo {year}
  {2018}{\natexlab{a}})}\BibitemShut {NoStop}%
\bibitem [{\citenamefont {Nair}\ \emph
  {et~al.}(2018{\natexlab{b}})\citenamefont {Nair}, \citenamefont {Ruf},
  \citenamefont {Schreiber}, \citenamefont {Miao}, \citenamefont {Grandon},
  \citenamefont {Baek}, \citenamefont {Goodge}, \citenamefont {Ruff},
  \citenamefont {Kourkoutis}, \citenamefont {Shen},\ and\ \citenamefont
  {Schlom}}]{Nair2018-214}%
  \BibitemOpen
  \bibfield  {author} {\bibinfo {author} {\bibfnamefont {H.~P.}\ \bibnamefont
  {Nair}}, \bibinfo {author} {\bibfnamefont {J.~P.}\ \bibnamefont {Ruf}},
  \bibinfo {author} {\bibfnamefont {N.~J.}\ \bibnamefont {Schreiber}}, \bibinfo
  {author} {\bibfnamefont {L.}~\bibnamefont {Miao}}, \bibinfo {author}
  {\bibfnamefont {M.~L.}\ \bibnamefont {Grandon}}, \bibinfo {author}
  {\bibfnamefont {D.~J.}\ \bibnamefont {Baek}}, \bibinfo {author}
  {\bibfnamefont {B.~H.}\ \bibnamefont {Goodge}}, \bibinfo {author}
  {\bibfnamefont {J.~P.~C.}\ \bibnamefont {Ruff}}, \bibinfo {author}
  {\bibfnamefont {L.~F.}\ \bibnamefont {Kourkoutis}}, \bibinfo {author}
  {\bibfnamefont {K.~M.}\ \bibnamefont {Shen}},\ and\ \bibinfo {author}
  {\bibfnamefont {D.~G.}\ \bibnamefont {Schlom}},\ }\href
  {https://doi.org/10.1063/1.5053084} {\bibfield  {journal} {\bibinfo
  {journal} {APL Materials}\ }\textbf {\bibinfo {volume} {6}},\ \bibinfo
  {pages} {101108} (\bibinfo {year} {2018}{\natexlab{b}})}\BibitemShut
  {NoStop}%
\bibitem [{\citenamefont {Marshall}\ \emph {et~al.}(2017)\citenamefont
  {Marshall}, \citenamefont {Kim}, \citenamefont {Ahadi},\ and\ \citenamefont
  {Stemmer}}]{Marshall2017}%
  \BibitemOpen
  \bibfield  {author} {\bibinfo {author} {\bibfnamefont {P.~B.}\ \bibnamefont
  {Marshall}}, \bibinfo {author} {\bibfnamefont {H.}~\bibnamefont {Kim}},
  \bibinfo {author} {\bibfnamefont {K.}~\bibnamefont {Ahadi}},\ and\ \bibinfo
  {author} {\bibfnamefont {S.}~\bibnamefont {Stemmer}},\ }\href
  {https://doi.org/10.1063/1.4998772} {\bibfield  {journal} {\bibinfo
  {journal} {APL Materials}\ }\textbf {\bibinfo {volume} {5}},\ \bibinfo
  {pages} {096101} (\bibinfo {year} {2017})}\BibitemShut {NoStop}%
\bibitem [{\citenamefont {Cao}\ \emph {et~al.}(2016)\citenamefont {Cao},
  \citenamefont {Massarotti}, \citenamefont {Vickers}, \citenamefont
  {Kursumovic}, \citenamefont {Bernardo}, \citenamefont {Robinson},
  \citenamefont {Tafuri}, \citenamefont {MacManus-Driscoll},\ and\
  \citenamefont {Blamire}}]{Cao2016}%
  \BibitemOpen
  \bibfield  {author} {\bibinfo {author} {\bibfnamefont {J.}~\bibnamefont
  {Cao}}, \bibinfo {author} {\bibfnamefont {D.}~\bibnamefont {Massarotti}},
  \bibinfo {author} {\bibfnamefont {M.~E.}\ \bibnamefont {Vickers}}, \bibinfo
  {author} {\bibfnamefont {A.}~\bibnamefont {Kursumovic}}, \bibinfo {author}
  {\bibfnamefont {A.~D.}\ \bibnamefont {Bernardo}}, \bibinfo {author}
  {\bibfnamefont {J.~W.~A.}\ \bibnamefont {Robinson}}, \bibinfo {author}
  {\bibfnamefont {F.}~\bibnamefont {Tafuri}}, \bibinfo {author} {\bibfnamefont
  {J.~L.}\ \bibnamefont {MacManus-Driscoll}},\ and\ \bibinfo {author}
  {\bibfnamefont {M.~G.}\ \bibnamefont {Blamire}},\ }\href
  {https://doi.org/10.1088/0953-2048/29/9/095005} {\bibfield  {journal}
  {\bibinfo  {journal} {Supercond. Sci. Technol.}\ }\textbf {\bibinfo {volume}
  {29}},\ \bibinfo {pages} {095005} (\bibinfo {year} {2016})}\BibitemShut
  {NoStop}%
\bibitem [{\citenamefont {Uchida}\ \emph {et~al.}(2017)\citenamefont {Uchida},
  \citenamefont {Ide}, \citenamefont {Watanabe}, \citenamefont {Takahashi},
  \citenamefont {Tokura},\ and\ \citenamefont {Kawasaki}}]{Uchida2017}%
  \BibitemOpen
  \bibfield  {author} {\bibinfo {author} {\bibfnamefont {M.}~\bibnamefont
  {Uchida}}, \bibinfo {author} {\bibfnamefont {M.}~\bibnamefont {Ide}},
  \bibinfo {author} {\bibfnamefont {H.}~\bibnamefont {Watanabe}}, \bibinfo
  {author} {\bibfnamefont {K.~S.}\ \bibnamefont {Takahashi}}, \bibinfo {author}
  {\bibfnamefont {Y.}~\bibnamefont {Tokura}},\ and\ \bibinfo {author}
  {\bibfnamefont {M.}~\bibnamefont {Kawasaki}},\ }\href
  {https://doi.org/10.1063/1.5007342} {\bibfield  {journal} {\bibinfo
  {journal} {APL Materials}\ }\textbf {\bibinfo {volume} {5}},\ \bibinfo
  {pages} {106108} (\bibinfo {year} {2017})}\BibitemShut {NoStop}%
\bibitem [{\citenamefont {Krockenberger}\ \emph {et~al.}(2010)\citenamefont
  {Krockenberger}, \citenamefont {Uchida}, \citenamefont {Takahashi},
  \citenamefont {Nakamura}, \citenamefont {Kawasaki},\ and\ \citenamefont
  {Tokura}}]{Krockenberger2010}%
  \BibitemOpen
  \bibfield  {author} {\bibinfo {author} {\bibfnamefont {Y.}~\bibnamefont
  {Krockenberger}}, \bibinfo {author} {\bibfnamefont {M.}~\bibnamefont
  {Uchida}}, \bibinfo {author} {\bibfnamefont {K.~S.}\ \bibnamefont
  {Takahashi}}, \bibinfo {author} {\bibfnamefont {M.}~\bibnamefont {Nakamura}},
  \bibinfo {author} {\bibfnamefont {M.}~\bibnamefont {Kawasaki}},\ and\
  \bibinfo {author} {\bibfnamefont {Y.}~\bibnamefont {Tokura}},\ }\href
  {https://doi.org/10.1063/1.3481363} {\bibfield  {journal} {\bibinfo
  {journal} {Appl. Phys. Lett.}\ }\textbf {\bibinfo {volume} {97}},\ \bibinfo
  {pages} {082502} (\bibinfo {year} {2010})}\BibitemShut {NoStop}%
\bibitem [{\citenamefont {Siemons}\ \emph {et~al.}(2007)\citenamefont
  {Siemons}, \citenamefont {Koster}, \citenamefont {Vailionis}, \citenamefont
  {Yamamoto}, \citenamefont {Blank},\ and\ \citenamefont
  {Beasley}}]{Siemons2007}%
  \BibitemOpen
  \bibfield  {author} {\bibinfo {author} {\bibfnamefont {W.}~\bibnamefont
  {Siemons}}, \bibinfo {author} {\bibfnamefont {G.}~\bibnamefont {Koster}},
  \bibinfo {author} {\bibfnamefont {A.}~\bibnamefont {Vailionis}}, \bibinfo
  {author} {\bibfnamefont {H.}~\bibnamefont {Yamamoto}}, \bibinfo {author}
  {\bibfnamefont {D.~H.~A.}\ \bibnamefont {Blank}},\ and\ \bibinfo {author}
  {\bibfnamefont {M.~R.}\ \bibnamefont {Beasley}},\ }\href
  {https://doi.org/10.1103/PhysRevB.76.075126} {\bibfield  {journal} {\bibinfo
  {journal} {Phys. Rev. B}\ }\textbf {\bibinfo {volume} {76}},\ \bibinfo
  {pages} {075126} (\bibinfo {year} {2007})}\BibitemShut {NoStop}%
\bibitem [{\citenamefont {Dabrowski}\ \emph {et~al.}(2004)\citenamefont
  {Dabrowski}, \citenamefont {Chmaissem}, \citenamefont {Klamut}, \citenamefont
  {Kolesnik}, \citenamefont {Maxwell}, \citenamefont {Mais}, \citenamefont
  {Ito}, \citenamefont {Armstrong}, \citenamefont {Jorgensen},\ and\
  \citenamefont {Short}}]{Dabrowski2004}%
  \BibitemOpen
  \bibfield  {author} {\bibinfo {author} {\bibfnamefont {B.}~\bibnamefont
  {Dabrowski}}, \bibinfo {author} {\bibfnamefont {O.}~\bibnamefont
  {Chmaissem}}, \bibinfo {author} {\bibfnamefont {P.~W.}\ \bibnamefont
  {Klamut}}, \bibinfo {author} {\bibfnamefont {S.}~\bibnamefont {Kolesnik}},
  \bibinfo {author} {\bibfnamefont {M.}~\bibnamefont {Maxwell}}, \bibinfo
  {author} {\bibfnamefont {J.}~\bibnamefont {Mais}}, \bibinfo {author}
  {\bibfnamefont {Y.}~\bibnamefont {Ito}}, \bibinfo {author} {\bibfnamefont
  {B.~D.}\ \bibnamefont {Armstrong}}, \bibinfo {author} {\bibfnamefont {J.~D.}\
  \bibnamefont {Jorgensen}},\ and\ \bibinfo {author} {\bibfnamefont
  {S.}~\bibnamefont {Short}},\ }\href
  {https://doi.org/10.1103/PhysRevB.70.014423} {\bibfield  {journal} {\bibinfo
  {journal} {Phys. Rev. B}\ }\textbf {\bibinfo {volume} {70}},\ \bibinfo
  {pages} {014423} (\bibinfo {year} {2004})}\BibitemShut {NoStop}%
\bibitem [{\citenamefont {Schraknepper}\ \emph {et~al.}(2015)\citenamefont
  {Schraknepper}, \citenamefont {Bäumer}, \citenamefont {Dittmann},\ and\
  \citenamefont {De~Souza}}]{Schraknepper2015}%
  \BibitemOpen
  \bibfield  {author} {\bibinfo {author} {\bibfnamefont {H.}~\bibnamefont
  {Schraknepper}}, \bibinfo {author} {\bibfnamefont {C.}~\bibnamefont
  {Bäumer}}, \bibinfo {author} {\bibfnamefont {R.}~\bibnamefont {Dittmann}},\
  and\ \bibinfo {author} {\bibfnamefont {R.~A.}\ \bibnamefont {De~Souza}},\
  }\href@noop {} {\bibfield  {journal} {\bibinfo  {journal} {Phys. Chem. Chem.
  Phys.}\ }\textbf {\bibinfo {volume} {17}},\ \bibinfo {pages} {1060} (\bibinfo
  {year} {2015})}\BibitemShut {NoStop}%
\bibitem [{\citenamefont {Schraknepper}\ \emph {et~al.}(2016)\citenamefont
  {Schraknepper}, \citenamefont {Bäumer}, \citenamefont {Gunkel},
  \citenamefont {Dittmann},\ and\ \citenamefont {De~Souza}}]{Schraknepper2016}%
  \BibitemOpen
  \bibfield  {author} {\bibinfo {author} {\bibfnamefont {H.}~\bibnamefont
  {Schraknepper}}, \bibinfo {author} {\bibfnamefont {C.}~\bibnamefont
  {Bäumer}}, \bibinfo {author} {\bibfnamefont {F.}~\bibnamefont {Gunkel}},
  \bibinfo {author} {\bibfnamefont {R.}~\bibnamefont {Dittmann}},\ and\
  \bibinfo {author} {\bibfnamefont {R.~A.}\ \bibnamefont {De~Souza}},\ }\href
  {https://doi.org/10.1063/1.4972996} {\bibfield  {journal} {\bibinfo
  {journal} {APL Materials}\ }\textbf {\bibinfo {volume} {4}},\ \bibinfo
  {pages} {126109} (\bibinfo {year} {2016})}\BibitemShut {NoStop}%
\bibitem [{\citenamefont {Mao}\ \emph {et~al.}(1999)\citenamefont {Mao},
  \citenamefont {Mori},\ and\ \citenamefont {Maeno}}]{Mao1999}%
  \BibitemOpen
  \bibfield  {author} {\bibinfo {author} {\bibfnamefont {Z.~Q.}\ \bibnamefont
  {Mao}}, \bibinfo {author} {\bibfnamefont {Y.}~\bibnamefont {Mori}},\ and\
  \bibinfo {author} {\bibfnamefont {Y.}~\bibnamefont {Maeno}},\ }\href
  {https://doi.org/10.1103/PhysRevB.60.610} {\bibfield  {journal} {\bibinfo
  {journal} {Phys. Rev. B}\ }\textbf {\bibinfo {volume} {60}},\ \bibinfo
  {pages} {610} (\bibinfo {year} {1999})}\BibitemShut {NoStop}%
\bibitem [{\citenamefont {Mackenzie}\ \emph {et~al.}(1998)\citenamefont
  {Mackenzie}, \citenamefont {Haselwimmer}, \citenamefont {Tyler},
  \citenamefont {Lonzarich}, \citenamefont {Mori}, \citenamefont {Nishizaki},\
  and\ \citenamefont {Maeno}}]{Mackenzie1998}%
  \BibitemOpen
  \bibfield  {author} {\bibinfo {author} {\bibfnamefont {A.~P.}\ \bibnamefont
  {Mackenzie}}, \bibinfo {author} {\bibfnamefont {R.~K.~W.}\ \bibnamefont
  {Haselwimmer}}, \bibinfo {author} {\bibfnamefont {A.~W.}\ \bibnamefont
  {Tyler}}, \bibinfo {author} {\bibfnamefont {G.~G.}\ \bibnamefont
  {Lonzarich}}, \bibinfo {author} {\bibfnamefont {Y.}~\bibnamefont {Mori}},
  \bibinfo {author} {\bibfnamefont {S.}~\bibnamefont {Nishizaki}},\ and\
  \bibinfo {author} {\bibfnamefont {Y.}~\bibnamefont {Maeno}},\ }\href
  {https://doi.org/10.1103/PhysRevLett.80.161} {\bibfield  {journal} {\bibinfo
  {journal} {Phys. Rev. Lett.}\ }\textbf {\bibinfo {volume} {80}},\ \bibinfo
  {pages} {161} (\bibinfo {year} {1998})}\BibitemShut {NoStop}%
\bibitem [{\citenamefont {Dietl}\ \emph {et~al.}(2018)\citenamefont {Dietl},
  \citenamefont {Sinha}, \citenamefont {Christiani}, \citenamefont {Khaydukov},
  \citenamefont {Keller}, \citenamefont {Putzky}, \citenamefont {Ibrahimkutty},
  \citenamefont {Wochner}, \citenamefont {Logvenov}, \citenamefont {van Aken},
  \citenamefont {Kim},\ and\ \citenamefont {Keimer}}]{Dietl2018}%
  \BibitemOpen
  \bibfield  {author} {\bibinfo {author} {\bibfnamefont {C.}~\bibnamefont
  {Dietl}}, \bibinfo {author} {\bibfnamefont {S.~K.}\ \bibnamefont {Sinha}},
  \bibinfo {author} {\bibfnamefont {G.}~\bibnamefont {Christiani}}, \bibinfo
  {author} {\bibfnamefont {Y.}~\bibnamefont {Khaydukov}}, \bibinfo {author}
  {\bibfnamefont {T.}~\bibnamefont {Keller}}, \bibinfo {author} {\bibfnamefont
  {D.}~\bibnamefont {Putzky}}, \bibinfo {author} {\bibfnamefont
  {S.}~\bibnamefont {Ibrahimkutty}}, \bibinfo {author} {\bibfnamefont
  {P.}~\bibnamefont {Wochner}}, \bibinfo {author} {\bibfnamefont
  {G.}~\bibnamefont {Logvenov}}, \bibinfo {author} {\bibfnamefont {P.~A.}\
  \bibnamefont {van Aken}}, \bibinfo {author} {\bibfnamefont {B.~J.}\
  \bibnamefont {Kim}},\ and\ \bibinfo {author} {\bibfnamefont {B.}~\bibnamefont
  {Keimer}},\ }\href {https://doi.org/10.1063/1.5007680} {\bibfield  {journal}
  {\bibinfo  {journal} {Appl. Phys. Lett.}\ }\textbf {\bibinfo {volume}
  {112}},\ \bibinfo {pages} {031902} (\bibinfo {year} {2018})}\BibitemShut
  {NoStop}%
\bibitem [{\citenamefont {Barla}\ \emph {et~al.}(2016)\citenamefont {Barla},
  \citenamefont {Nicol{\'{a}}s}, \citenamefont {Cocco}, \citenamefont
  {Valvidares}, \citenamefont {Herrero-Mart{\'\i}n}, \citenamefont {Gargiani},
  \citenamefont {Moldes}, \citenamefont {Ruget}, \citenamefont {Pellegrin},\
  and\ \citenamefont {Ferrer}}]{Barla2016}%
  \BibitemOpen
  \bibfield  {author} {\bibinfo {author} {\bibfnamefont {A.}~\bibnamefont
  {Barla}}, \bibinfo {author} {\bibfnamefont {J.}~\bibnamefont
  {Nicol{\'{a}}s}}, \bibinfo {author} {\bibfnamefont {D.}~\bibnamefont
  {Cocco}}, \bibinfo {author} {\bibfnamefont {S.~M.}\ \bibnamefont
  {Valvidares}}, \bibinfo {author} {\bibfnamefont {J.}~\bibnamefont
  {Herrero-Mart{\'\i}n}}, \bibinfo {author} {\bibfnamefont {P.}~\bibnamefont
  {Gargiani}}, \bibinfo {author} {\bibfnamefont {J.}~\bibnamefont {Moldes}},
  \bibinfo {author} {\bibfnamefont {C.}~\bibnamefont {Ruget}}, \bibinfo
  {author} {\bibfnamefont {E.}~\bibnamefont {Pellegrin}},\ and\ \bibinfo
  {author} {\bibfnamefont {S.}~\bibnamefont {Ferrer}},\ }\href@noop {}
  {\bibfield  {journal} {\bibinfo  {journal} {Journal of Synchrotron
  Radiation}\ }\textbf {\bibinfo {volume} {23}},\ \bibinfo {pages} {1507}
  (\bibinfo {year} {2016})}\BibitemShut {NoStop}%
\bibitem [{\citenamefont {Nobukane}\ \emph {et~al.}(2017)\citenamefont
  {Nobukane}, \citenamefont {Matsuyama},\ and\ \citenamefont
  {Tanda}}]{Nobukane2017}%
  \BibitemOpen
  \bibfield  {author} {\bibinfo {author} {\bibfnamefont {H.}~\bibnamefont
  {Nobukane}}, \bibinfo {author} {\bibfnamefont {T.}~\bibnamefont
  {Matsuyama}},\ and\ \bibinfo {author} {\bibfnamefont {S.}~\bibnamefont
  {Tanda}},\ }\href@noop {} {\bibfield  {journal} {\bibinfo  {journal} {Sci.
  Rep.}\ }\textbf {\bibinfo {volume} {7}},\ \bibinfo {pages} {41291} (\bibinfo
  {year} {2017})}\BibitemShut {NoStop}%
\bibitem [{\citenamefont {Schmidt}\ \emph {et~al.}(1996)\citenamefont
  {Schmidt}, \citenamefont {Cummins}, \citenamefont {B\"urk}, \citenamefont
  {Lu}, \citenamefont {N\"ucker}, \citenamefont {Schuppler},\ and\
  \citenamefont {Lichtenberg}}]{Schmidt1996}%
  \BibitemOpen
  \bibfield  {author} {\bibinfo {author} {\bibfnamefont {M.}~\bibnamefont
  {Schmidt}}, \bibinfo {author} {\bibfnamefont {T.~R.}\ \bibnamefont
  {Cummins}}, \bibinfo {author} {\bibfnamefont {M.}~\bibnamefont {B\"urk}},
  \bibinfo {author} {\bibfnamefont {D.~H.}\ \bibnamefont {Lu}}, \bibinfo
  {author} {\bibfnamefont {N.}~\bibnamefont {N\"ucker}}, \bibinfo {author}
  {\bibfnamefont {S.}~\bibnamefont {Schuppler}},\ and\ \bibinfo {author}
  {\bibfnamefont {F.}~\bibnamefont {Lichtenberg}},\ }\href
  {https://doi.org/10.1103/PhysRevB.53.R14761} {\bibfield  {journal} {\bibinfo
  {journal} {Phys. Rev. B}\ }\textbf {\bibinfo {volume} {53}},\ \bibinfo
  {pages} {R14761} (\bibinfo {year} {1996})}\BibitemShut {NoStop}%
\bibitem [{\citenamefont {Moon}\ \emph {et~al.}(2006)\citenamefont {Moon},
  \citenamefont {Kim}, \citenamefont {Kim}, \citenamefont {Lee}, \citenamefont
  {Kim}, \citenamefont {Park}, \citenamefont {Kim}, \citenamefont {Oh},
  \citenamefont {Nakatsuji}, \citenamefont {Maeno}, \citenamefont {Nagai},
  \citenamefont {Ikeda}, \citenamefont {Cao},\ and\ \citenamefont
  {Noh}}]{Moon2006}%
  \BibitemOpen
  \bibfield  {author} {\bibinfo {author} {\bibfnamefont {S.~J.}\ \bibnamefont
  {Moon}}, \bibinfo {author} {\bibfnamefont {M.~W.}\ \bibnamefont {Kim}},
  \bibinfo {author} {\bibfnamefont {K.~W.}\ \bibnamefont {Kim}}, \bibinfo
  {author} {\bibfnamefont {Y.~S.}\ \bibnamefont {Lee}}, \bibinfo {author}
  {\bibfnamefont {J.-Y.}\ \bibnamefont {Kim}}, \bibinfo {author} {\bibfnamefont
  {J.-H.}\ \bibnamefont {Park}}, \bibinfo {author} {\bibfnamefont {B.~J.}\
  \bibnamefont {Kim}}, \bibinfo {author} {\bibfnamefont {S.-J.}\ \bibnamefont
  {Oh}}, \bibinfo {author} {\bibfnamefont {S.}~\bibnamefont {Nakatsuji}},
  \bibinfo {author} {\bibfnamefont {Y.}~\bibnamefont {Maeno}}, \bibinfo
  {author} {\bibfnamefont {I.}~\bibnamefont {Nagai}}, \bibinfo {author}
  {\bibfnamefont {S.~I.}\ \bibnamefont {Ikeda}}, \bibinfo {author}
  {\bibfnamefont {G.}~\bibnamefont {Cao}},\ and\ \bibinfo {author}
  {\bibfnamefont {T.~W.}\ \bibnamefont {Noh}},\ }\href
  {https://doi.org/10.1103/PhysRevB.74.113104} {\bibfield  {journal} {\bibinfo
  {journal} {Phys. Rev. B}\ }\textbf {\bibinfo {volume} {74}},\ \bibinfo
  {pages} {113104} (\bibinfo {year} {2006})}\BibitemShut {NoStop}%
\bibitem [{\citenamefont {Pchelkina}\ \emph {et~al.}(2007)\citenamefont
  {Pchelkina}, \citenamefont {Nekrasov}, \citenamefont {Pruschke},
  \citenamefont {Sekiyama}, \citenamefont {Suga}, \citenamefont {Anisimov},\
  and\ \citenamefont {Vollhardt}}]{Pchelkina2007}%
  \BibitemOpen
  \bibfield  {author} {\bibinfo {author} {\bibfnamefont {Z.~V.}\ \bibnamefont
  {Pchelkina}}, \bibinfo {author} {\bibfnamefont {I.~A.}\ \bibnamefont
  {Nekrasov}}, \bibinfo {author} {\bibfnamefont {T.}~\bibnamefont {Pruschke}},
  \bibinfo {author} {\bibfnamefont {A.}~\bibnamefont {Sekiyama}}, \bibinfo
  {author} {\bibfnamefont {S.}~\bibnamefont {Suga}}, \bibinfo {author}
  {\bibfnamefont {V.~I.}\ \bibnamefont {Anisimov}},\ and\ \bibinfo {author}
  {\bibfnamefont {D.}~\bibnamefont {Vollhardt}},\ }\href
  {https://doi.org/10.1103/PhysRevB.75.035122} {\bibfield  {journal} {\bibinfo
  {journal} {Phys. Rev. B}\ }\textbf {\bibinfo {volume} {75}},\ \bibinfo
  {pages} {035122} (\bibinfo {year} {2007})}\BibitemShut {NoStop}%
\bibitem [{\citenamefont {Iliev}\ \emph {et~al.}(2005)\citenamefont {Iliev},
  \citenamefont {Popov}, \citenamefont {Litvinchuk}, \citenamefont {Abrashev},
  \citenamefont {Bäckström}, \citenamefont {Sun}, \citenamefont {Meng},\ and\
  \citenamefont {Chu}}]{Iliev2005}%
  \BibitemOpen
  \bibfield  {author} {\bibinfo {author} {\bibfnamefont {M.~N.}\ \bibnamefont
  {Iliev}}, \bibinfo {author} {\bibfnamefont {V.~N.}\ \bibnamefont {Popov}},
  \bibinfo {author} {\bibfnamefont {A.~P.}\ \bibnamefont {Litvinchuk}},
  \bibinfo {author} {\bibfnamefont {M.~V.}\ \bibnamefont {Abrashev}}, \bibinfo
  {author} {\bibfnamefont {J.}~\bibnamefont {Bäckström}}, \bibinfo {author}
  {\bibfnamefont {Y.~Y.}\ \bibnamefont {Sun}}, \bibinfo {author} {\bibfnamefont
  {R.~L.}\ \bibnamefont {Meng}},\ and\ \bibinfo {author} {\bibfnamefont
  {C.~W.}\ \bibnamefont {Chu}},\ }\href@noop {} {\bibfield  {journal} {\bibinfo
   {journal} {Physica B: Condensed Matter}\ }\textbf {\bibinfo {volume}
  {358}},\ \bibinfo {pages} {138 } (\bibinfo {year} {2005})}\BibitemShut
  {NoStop}%
\bibitem [{\citenamefont {Hepting}\ \emph {et~al.}(2014)\citenamefont
  {Hepting}, \citenamefont {Minola}, \citenamefont {Frano}, \citenamefont
  {Cristiani}, \citenamefont {Logvenov}, \citenamefont {Schierle},
  \citenamefont {Wu}, \citenamefont {Bluschke}, \citenamefont {Weschke},
  \citenamefont {Habermeier}, \citenamefont {Benckiser}, \citenamefont
  {Le~Tacon},\ and\ \citenamefont {Keimer}}]{Hepting2014}%
  \BibitemOpen
  \bibfield  {author} {\bibinfo {author} {\bibfnamefont {M.}~\bibnamefont
  {Hepting}}, \bibinfo {author} {\bibfnamefont {M.}~\bibnamefont {Minola}},
  \bibinfo {author} {\bibfnamefont {A.}~\bibnamefont {Frano}}, \bibinfo
  {author} {\bibfnamefont {G.}~\bibnamefont {Cristiani}}, \bibinfo {author}
  {\bibfnamefont {G.}~\bibnamefont {Logvenov}}, \bibinfo {author}
  {\bibfnamefont {E.}~\bibnamefont {Schierle}}, \bibinfo {author}
  {\bibfnamefont {M.}~\bibnamefont {Wu}}, \bibinfo {author} {\bibfnamefont
  {M.}~\bibnamefont {Bluschke}}, \bibinfo {author} {\bibfnamefont
  {E.}~\bibnamefont {Weschke}}, \bibinfo {author} {\bibfnamefont {H.-U.}\
  \bibnamefont {Habermeier}}, \bibinfo {author} {\bibfnamefont
  {E.}~\bibnamefont {Benckiser}}, \bibinfo {author} {\bibfnamefont
  {M.}~\bibnamefont {Le~Tacon}},\ and\ \bibinfo {author} {\bibfnamefont
  {B.}~\bibnamefont {Keimer}},\ }\href
  {https://doi.org/10.1103/PhysRevLett.113.227206} {\bibfield  {journal}
  {\bibinfo  {journal} {Phys. Rev. Lett.}\ }\textbf {\bibinfo {volume} {113}},\
  \bibinfo {pages} {227206} (\bibinfo {year} {2014})}\BibitemShut {NoStop}%
\bibitem [{\citenamefont {Kim}\ \emph {et~al.}(2017)\citenamefont {Kim},
  \citenamefont {Christiani}, \citenamefont {Logvenov}, \citenamefont {Choi},
  \citenamefont {Kim}, \citenamefont {Minola},\ and\ \citenamefont
  {Keimer}}]{Kim2017}%
  \BibitemOpen
  \bibfield  {author} {\bibinfo {author} {\bibfnamefont {G.}~\bibnamefont
  {Kim}}, \bibinfo {author} {\bibfnamefont {G.}~\bibnamefont {Christiani}},
  \bibinfo {author} {\bibfnamefont {G.}~\bibnamefont {Logvenov}}, \bibinfo
  {author} {\bibfnamefont {S.}~\bibnamefont {Choi}}, \bibinfo {author}
  {\bibfnamefont {H.-H.}\ \bibnamefont {Kim}}, \bibinfo {author} {\bibfnamefont
  {M.}~\bibnamefont {Minola}},\ and\ \bibinfo {author} {\bibfnamefont
  {B.}~\bibnamefont {Keimer}},\ }\href
  {https://doi.org/10.1103/PhysRevMaterials.1.054801} {\bibfield  {journal}
  {\bibinfo  {journal} {Phys. Rev. Materials}\ }\textbf {\bibinfo {volume}
  {1}},\ \bibinfo {pages} {054801} (\bibinfo {year} {2017})}\BibitemShut
  {NoStop}%
\bibitem [{\citenamefont {Sakita}\ \emph {et~al.}(2001)\citenamefont {Sakita},
  \citenamefont {Nimori}, \citenamefont {Mao}, \citenamefont {Maeno},
  \citenamefont {Ogita},\ and\ \citenamefont {Udagawa}}]{Sakita2001}%
  \BibitemOpen
  \bibfield  {author} {\bibinfo {author} {\bibfnamefont {S.}~\bibnamefont
  {Sakita}}, \bibinfo {author} {\bibfnamefont {S.}~\bibnamefont {Nimori}},
  \bibinfo {author} {\bibfnamefont {Z.~Q.}\ \bibnamefont {Mao}}, \bibinfo
  {author} {\bibfnamefont {Y.}~\bibnamefont {Maeno}}, \bibinfo {author}
  {\bibfnamefont {N.}~\bibnamefont {Ogita}},\ and\ \bibinfo {author}
  {\bibfnamefont {M.}~\bibnamefont {Udagawa}},\ }\href
  {https://doi.org/10.1103/PhysRevB.63.134520} {\bibfield  {journal} {\bibinfo
  {journal} {Phys. Rev. B}\ }\textbf {\bibinfo {volume} {63}},\ \bibinfo
  {pages} {134520} (\bibinfo {year} {2001})}\BibitemShut {NoStop}%
\bibitem [{\citenamefont {Braden}\ \emph {et~al.}(2007)\citenamefont {Braden},
  \citenamefont {Reichardt}, \citenamefont {Sidis}, \citenamefont {Mao},\ and\
  \citenamefont {Maeno}}]{Braden2007}%
  \BibitemOpen
  \bibfield  {author} {\bibinfo {author} {\bibfnamefont {M.}~\bibnamefont
  {Braden}}, \bibinfo {author} {\bibfnamefont {W.}~\bibnamefont {Reichardt}},
  \bibinfo {author} {\bibfnamefont {Y.}~\bibnamefont {Sidis}}, \bibinfo
  {author} {\bibfnamefont {Z.}~\bibnamefont {Mao}},\ and\ \bibinfo {author}
  {\bibfnamefont {Y.}~\bibnamefont {Maeno}},\ }\href@noop {} {\bibfield
  {journal} {\bibinfo  {journal} {Phys. Rev. B}\ }\textbf {\bibinfo {volume}
  {76}},\ \bibinfo {pages} {014505} (\bibinfo {year} {2007})}\BibitemShut
  {NoStop}%
\bibitem [{\citenamefont {Ammundsen}\ \emph {et~al.}(1999)\citenamefont
  {Ammundsen}, \citenamefont {Burns}, \citenamefont {Islam}, \citenamefont
  {Kanoh},\ and\ \citenamefont {Rozière}}]{Ammundsen1999}%
  \BibitemOpen
  \bibfield  {author} {\bibinfo {author} {\bibfnamefont {B.}~\bibnamefont
  {Ammundsen}}, \bibinfo {author} {\bibfnamefont {G.~R.}\ \bibnamefont
  {Burns}}, \bibinfo {author} {\bibfnamefont {M.~S.}\ \bibnamefont {Islam}},
  \bibinfo {author} {\bibfnamefont {H.}~\bibnamefont {Kanoh}},\ and\ \bibinfo
  {author} {\bibfnamefont {J.}~\bibnamefont {Rozière}},\ }\href
  {https://doi.org/10.1021/jp984398l} {\bibfield  {journal} {\bibinfo
  {journal} {J. Phys. Chem. B}\ }\textbf {\bibinfo {volume} {103}},\ \bibinfo
  {pages} {5175} (\bibinfo {year} {1999})}\BibitemShut {NoStop}%
\bibitem [{\citenamefont {Limmer}\ \emph {et~al.}(1998)\citenamefont {Limmer},
  \citenamefont {Ritter}, \citenamefont {Sauer}, \citenamefont {Mensching},
  \citenamefont {Liu},\ and\ \citenamefont {Rauschenbach}}]{Limmer1998}%
  \BibitemOpen
  \bibfield  {author} {\bibinfo {author} {\bibfnamefont {W.}~\bibnamefont
  {Limmer}}, \bibinfo {author} {\bibfnamefont {W.}~\bibnamefont {Ritter}},
  \bibinfo {author} {\bibfnamefont {R.}~\bibnamefont {Sauer}}, \bibinfo
  {author} {\bibfnamefont {B.}~\bibnamefont {Mensching}}, \bibinfo {author}
  {\bibfnamefont {C.}~\bibnamefont {Liu}},\ and\ \bibinfo {author}
  {\bibfnamefont {B.}~\bibnamefont {Rauschenbach}},\ }\href
  {https://doi.org/10.1063/1.121426} {\bibfield  {journal} {\bibinfo  {journal}
  {Appl. Phys. Lett.}\ }\textbf {\bibinfo {volume} {72}},\ \bibinfo {pages}
  {2589} (\bibinfo {year} {1998})}\BibitemShut {NoStop}%
\end{thebibliography}%

\end{document}